\documentstyle[12pt]{article}
\newcommand{\bq}{\begin{equation}}
\newcommand{\eq}{\end{equation}}
\newcommand{\ba}{\begin{eqnarray}}
\newcommand{\ea}{\end{eqnarray}}

\newcommand{\req}[1]{(\ref{#1})}
\begin{document}
\newtheorem{tab}[table]{Table}
\voffset -2cm
\begin{flushleft}
LMU-09/97
\end{flushleft}
\begin{center}
\vspace{1.0cm}\hfill\\
{ \Large \bf Scaling of $Z'$ exclusion limits and $Z'$ measurements\\
with energy, luminosity and systematic errors}\footnote{Talk presented
at the joint meeting of the 
networks  "Fundamental Structure of Matter" and "Tests of the
Electroweak Symmetry Breaking", Ouranoupolis, Greece, May 1997.}
\vspace{0.5cm}\\ 
A. Leike
\\
{\it Ludwigs--Maximilians-Universit\"at, Sektion Physik, Theresienstr. 37,\\
D-80333 M\"unchen, Germany}\\
E-mail: leike@graviton.hep.physik.uni-muenchen.de\\
\end{center}
\hfill\vspace{1cm}\\
\noindent
{\small 
The influence of the c.m. energy, integrated luminosity and
systematic errors on $Z'$ exclusion limits and on errors of 
$Z'$ model measurements is discussed.  
Simple scaling laws are given for $e^+e^-,\ pp(p\bar p)$ and $e^\pm p$
colliders and compared with results of exact analyses.
}
 \vspace{0.5cm}
%
\section{Introduction} 
%
The present colliders at CERN, SLAC, Fermilab and DESY have tested the
Standard Model (SM) with a precision never reached before. 
The next generation of colliders will reach energies  at the parton
level, which are of the order of the
scale of the electroweak symmetry breaking.

There is a common belief and there are experimental hints that the SM
cannot be the ultimate theory of elementary particle physics.
Among many theoretical problems of the SM, a more general theory must
explain why the scale of the electroweak symmetry breaking is
seventeen orders of magnitude smaller than the Planck mass.
In a quantum theory explaining the electroweak mass scale, some
particles are expected to have a mass of this order.
Then they can be produced at the next generation of colliders.

Electroweak and strong interactions are described by gauge theories.
In grand unified theories (GUT's), all interactions are unified in one
simple gauge group at very high energies. 
GUT's, which are compatible with the present data, predict new particles.
The masses of these new particles depend on details of the
breaking of the large gauge symmetry.
An observation of particles associated with a GUT would
therefore provide information on the GUT group and
on its symmetry breaking.

Extra neutral gauge bosons ($Z'$) are predicted in GUT's with a unification
group larger than $SU(5)$.
They are candidates of particles with a mass of
the order of the electroweak symmetry breaking scale.
Therefore, the search for a $Z'$ is an important part of the
scientific program of every present and future collider.

In the previous years, extensive studies on the sensitivity of future
$e^+e^-$ \cite{ee}, $pp(p\bar p)$ \cite{tevatron} and $e^\pm p$
\cite{hera} colliders to a $Z'$ have been completed. 
See these references for more details and for a list
of related original references. 

In this contribution, I concentrate on the discussion of simple
analytic formulae, 
which show the main dependence of the $Z'$ bounds on the c.m. energy
$\sqrt{s}$, the integrated luminosity $L$ and the experimental error
$\Delta O$ of the observable $O$. 
The experimental error consists of a systematic
$\Delta^{syst} O$ and a statistical $\Delta^{stat}O$ contribution.
Both contributions are added in quadrature,
\bq
\Delta O = \sqrt{(\Delta^{stat}O)^2+(\Delta^{syst}O)^2}
= \Delta^{stat}O\cdot\sqrt{1+r^2},\ \ \ r=\Delta^{syst}O/\Delta^{stat}O
\eq 

One has to distinguish between $Z'$ {\it exclusion
limits} and $Z'$ {\it model measurements}.
Exclusion limits will be obtained if there are no
deviations from the SM.
All present bounds on $Z'$ theories are examples of exclusion limits.
Model measurements will be possible if there are deviations from the SM
predictions compatible with theories containing a $Z'$. 

To simplify the discussion, we
neglect here a possible mixing between the SM $Z$ boson and a $Z'$.
Furthermore, we assume no mixing between the SM fermions and the new
fermions present in a GUT.
We assume that all new particles in the GUT are heavier than the $Z'$.
The following discussion is qualitatively independent on these assumptions.
It is only weakly influenced by details of the $Z'$ model, which we
ignore in this contribution.
See \cite{e6} for a review on $Z'$ models and further references.
\section{$e^+e^-$ Colliders}
First signals of a $Z'$ could be observed in the reaction
$e^+e^-\rightarrow f\bar f$ at energies far below the $Z'$ peak.
A $Z'$ modifies cross sections and asymmetries due to 
interferences of the $Z'$ amplitude with the SM amplitudes.
Only the ratios $a'_f/M_{Z'}$ and $v'_f/M_{Z'}$ can be constrained
below the $Z'$ resonance and not the couplings and the mass separately. 

We do not consider here explicitely Bhabha and M{\o}ller scattering.
They give scalings similar to fermion pair production.
$W$ pair production is interesting in the case of $ZZ'$ mixing,
which we set to zero here.
\subsection{Exclusion Limits}
Assume that the SM $Z$ couples with a strength $g_1$ and
the $Z'$ with a strength $g_2$ to SM fermions.
We have $g_2/g_1=\sqrt{\frac{5}{3}}\sin\theta_W\approx 0.62$ in the
$E_6$ GUT. 
If future measurements of an observable $O$ agree with the SM
prediction, the $Z'$ must be heavy, 
\bq
\label{offres}
M_{Z'}>M_{Z'}^{lim}
\approx\frac{g_2}{g_1}\sqrt{s\frac{O}{\Delta O}}
\sim\left[\frac{sL}{1+r^2}\right]^{1/4}.
\eq
The scaling of the error with $s$ and $L$ ,
$\frac{\Delta O}{O}\approx\sqrt{(1+r^2)/N}\sim\sqrt{(1+r^2)s/L}$,
is taken into account in the last step of the estimate \req{offres}.
$M_{Z'}^{lim}$ depends on the fourth root of the experimental error.
The dependence on the systematic error will be suppressed if
it is not too large.
Suppose that an analysis gives certain exclusion limits $M_{Z'}^{lim}$
without systematic errors.
What changes are expected after the inclusion of systematic errors
assuming $r=1$?
The estimate \req{offres} predicts 
$M_{Z'}^{lim}\rightarrow M_{Z'}^{lim}/\sqrt[4]{2}$,
which is a reduction by 16\% only.

%
\begin{table}[tbh]
\begin{center}
\begin{tabular}{lrrrr}\hline
 &$\chi$ &$\psi$ &$\eta$ &$LR$ \rule[-2ex]{0ex}{5ex}\\ 
\hline
$M_{Z'}^{lim}$ \ stat. & 3.1 & 1.8 & 1.9 & 3.8 \\
$M_{Z'}^{lim}$ +syst. & 2.8 & 1.6 & 1.7 & 3.2 \\
$P_V^l$ \ stat. & 2.00$\pm$ 0.11 & 0.00$\pm$ 0.064 & -3.00$^{+0.53}_{-0.85}$ &
                         -0.148$^{+0.020}_{-0.024}$ \\
$P_V^l$ +syst.& 2.00$\pm$ 0.15 &
                          0.00$\pm$ 0.13 &
                         -3.00$^{+0.73}_{-1.55}$  & 
                         -0.148$^{+0.023}_{-0.026}$  \\
$P_L^b$ \ stat  & -0.500$\pm$ 0.018 &
                           0.500$\pm$ 0.035 &
                           2.00$^{+0.33}_{-0.31}$ &
                           -0.143$\pm$0.033  \\
$P_L^b$ +syst.& -0.500$\pm$ 0.070 &
                           0.500$\pm$ 0.130 &
                           2.00$^{+0.64}_{-0.62}$ &
                           -0.143$\pm$0.066  \\
\hline
\end{tabular}\medskip
\end{center}
{\small\it  \begin{tab}\label{zplimlc} The lower bound on $Z'$ masses
$M_{Z'}^{lim}$ in TeV excluded by $e^+e^-\rightarrow f\bar f$ at
$\sqrt{s}=0.5\,TeV$ and $L=20\,fb^{-1}$ (first two rows).
The $Z'$ coupling combinations $P_V^l=v'_l/a'_l$ and
$P_L^b=(v'_b+a'_b)/(2a'_b)$ and their 1-$\sigma$ errors (last four
rows) are derived under the same conditions as the exclusion 
limits but assuming $M_{Z'}=1\,TeV$.
The $\chi,\psi,\eta$ and $LR$ models are the same as in the Particle
Data Book.
The numbers are given with and without systematic errors.
They are taken from reference \cite{lmu0296}.
\end{tab}} \end{table}
%

Let us confront these findings with the numbers quoted in table~\ref{zplimlc}.
They include all SM corrections. 
The systematic errors included for observables with leptons in
the final state are roughly as large as their statistical
errors, i.e. $r\approx 1$. 
The systematic errors of observables with $b$ quarks in the final
state are roughly 4 times as large as the statistical errors,
i.e. $r\approx 4$.

We see that the predicted reduction of $M_{Z'}^{lim}$ by 16\% is
reproduced by the numbers in the first two rows of table~\ref{zplimlc}.
Although $M_{Z'}^{lim}$ is defined by hadronic and leptonic observables,
hadronic observables with large systematic errors don't spoil the
estimate \req{offres} because their contribution to $M_{Z'}^{lim}$
decreases in that case.
\subsection{Model Measurements}
We now assume that there exists a $Z'$ with $M_{Z'} <M_{Z'}^{lim}$.
Then, a measurement of $Z'$ model parameters is possible.
One can measure the $Z'$ mass for fixed couplings or the
coupling strength for a fixed $Z'$ mass (which could be known
from hadron collisions).
Considerations \cite{habil} similar to the previous section allow an
estimate of the {\it errors} of such measurements as
\bq
\label{epemmeas}
\frac{\Delta M_{Z'}}{M_{Z'}},\ \frac{\Delta g_2}{g_2} \approx
\frac{1}{2}\left[\frac{1+r^2}{sL}\right]^{1/2}.
\eq
We remark that the scalings \req{offres} and \req{epemmeas} depend on
the product $sL$ only.

Model measurements depend on the square root of systematic errors.
Another important difference to the exclusion limit \req{offres} is
that the couplings of the $Z'$ to leptons are measured by
observables with leptons in the final state only, while the 
couplings of the $Z'$ to $b$-quarks are measured by
observables with $b$-quarks in the final state only.

In particular, the estimate \req{epemmeas} predicts (under the
assumptions of the analysis \cite{lmu0296}) that the 
errors of measurements of the $Z'$ couplings to leptons ($b$-quarks)
change as $P_V^l\rightarrow P_V^l\sqrt{1+1^2}$ ($P_L^b\rightarrow
P_L^b\sqrt{1+4^2}$) after the inclusion of the systematic errors.
Of course, these predictions are only rough
approximations because they ignore details of the $Z'$ models and
differences of the ratio $\Delta^{syst}O/\Delta^{stat}O$ for the various
observables entering the analysis.
Nevertheless, they reproduce the main tendency of the last four rows in
table~\ref{zplimlc}. 
The estimate \req{epemmeas} explains why $Z'$ model measurements are
much more sensitive to systematic errors than $Z'$ exclusion limits.
\section{$pp$ or $p\bar p$ Colliders}
A $Z'$ can be produced on resonance in $pp$ or $p\bar p$ collisions.
The cleanest $Z'$ signature comes from the decay
$Z'\rightarrow\mu^+\mu^-$, which also gives the best present limits.
This signature has no SM background because the invariant energy of
the muon pair equals to the $Z'$ mass. 
We ignore here all other decay modes, although they are
important for model measurements.
\subsection{Exclusion Limits}
We approximate $\sigma(pp\rightarrow Z'\rightarrow \mu^+\mu^-)$
to derive \cite{leikezppp} an estimate for $M_{Z'}^{lim}$,
\bq
\label{fdef4}
M_{Z'}^{lim}\approx\frac{\sqrt{s}}{A}
\ln\left(\frac{L}{s}\frac{c_{Z'}C}{N_{Z'}}\right).
\eq
The constants $A$ and $C$ dependent only on the colliding particles,
$A=32\ (20),\ C=600\ (300)$ for ($pp$ or $p\bar p$) collisions. 
The constant $c_{Z'}$ depends on the $Z'$ model and on the colliding particles.
We have $c_{Z'}\approx 10^{-3}\ (0.5\cdot 10^{-3})$ for
the $E_6$ GUT in ($pp$ or $p\bar p$) collisions.
$N_{Z'}$ is the number of events corresponding to an exclusion limit
of a given confidence. 
For no observed events with no background, $N_{Z'}=3$ gives exclusion
limits with 95\% confidence. 

The estimate \req{fdef4} predicts
$M_{Z'}^{lim}=640\,GeV$ (95\% CL.) for $E_6$ GUT's at a $p\bar p$ 
collider with $\sqrt{s}=1.8\,TeV$ and $L=110pb^{-1}$. 
This can be confronted with the exclusion limits 
between $580$ and $620\,GeV$ obtained from CDF
under these experimental conditions \cite{cdf1196}.

As in the case of $e^+e^-$ collisions, the influence of systematic
errors can be estimated by a replacement of the luminosity,
$L\rightarrow L/(1+r^2)$.
The $Z'$ exclusion limit \req{fdef4} is 
almost unchanged by the inclusion of systematic errors because $L$
enters under the logarithm.
A decrease of $L$ by a factor 2 leads to a decrease of
$M_{Z'}^{lim}$ by only 9\% (7\%) in $pp$($p\bar p$) collisions.
\subsection{Model Measurements}
Suppose that there exists a $Z'$ with $M_{Z'} <M_{Z'}^{lim}$.
Assume that $N_{Z'}$ extra neutral gauge bosons are
detected.
The error of measurements of asymmetries $A_X$ and branching ratios
$Br_X$ at the $Z'$ peak can then be estimated as \cite{leikezppp}
\bq
\label{fdef5}
A_X,\ Br_X \approx\frac{1}{\sqrt{N_{Z'}}}
\approx \sqrt{\frac{s}{L}\frac{1}{c_{Z'}C}}
\exp\left\{\frac{AM_{Z'}}{2\sqrt{s}}\right\}.
\eq
$A_X$ and $Br_X$ define the couplings of the $Z'$ to fermions.
The relative errors of these measurements are given by the
estimate \req{fdef5} too.
In contrast to $e^+e^-$ collisions, $s$ and $L$ enter the scaling \req{fdef5}
non-symmetrically.
The dependence of model measurements on the integrated
luminosity, however, is the same as in $e^+e^-$ collisions.
A model measurement at hadron colliders is therefore as sensitive to
systematic errors as it is at $e^+e^-$ colliders.
In addition, the error of model measurements has an enhanced sensitivity
to errors of the c.m. energy $\sqrt{s}$.
\section{$e^\pm p$ Colliders}
A $Z'$ signal in $e^\pm p$ collisions is indirect as it is in $e^+e^-$
collisions.
A $Z'$ can only be observed through deviations of observables from
their SM prediction, which arise due to the interferences of the
additional $Z'$ contributions with the SM contributions.
\subsection{Exclusion Limits}
We get the scaling of $M_{Z'}^{lim}$ by considerations similar to
those explained for $e^+e^-$ collisions,
\bq
\label{zphera}
M_{Z'}^{lim}
\sim\left[\frac{Q^2L}{1+r^2}\right]^{1/4},
\eq
where $Q^2$ is the energy-momentum transferred in the $t$ channel.
The dependence on the luminosity is the same as in
$e^+e^-$ collisions. 
Therefore, the dependence on systematic errors is the same.
Unfortunately, the errors in $e^\pm p$ collisions are larger
than those in $e^+e^-$ collisions.
This is due to the small statistics at high
$Q^2$ and to the more complicated hadronic environment.
Therefore, $e^\pm p$ collisions are less sensitive to extra neutral
gauge bosons.

%
\begin{table}[tbh]
\begin{center}
\begin{tabular}{lrrrr}\hline
$M_{Z'}/GeV$ &$\chi$ &$\psi$ &$\eta$ &$LR$ \rule[-2ex]{0ex}{5ex}\\ 
\hline
$L=0.5\,fb^{-1}$ & 390 & 210 & 240 & 420 \\
$L=1.0\,fb^{-1}$ & 470 & 260 & 290 & 500 \\
ratio            &1.21 &1.24 &1.21 &1.19 \\
\hline
\end{tabular}\medskip
\end{center}
{\small\it  \begin{tab}\label{zplimhera} The 95\% CL predictions for
$M_{Z'}^{lim}$ from HERA with $\sqrt{s}=314\,GeV$ and the integrated
luminosities quoted in the table.
The first two rows are taken from table~3 of reference \cite{heralim}.
\end{tab}} \end{table}
%

The scaling \req{zphera} can be compared with the results of the
analysis \cite{heralim} quoted in table \ref{zplimhera}.
The ratio of the exclusion limits
is predicted to be $\sqrt[4]{2}\approx 1.19$, which
is in good agreement with the numbers in table~\ref{zplimhera}.
\subsection{Model Measurements}
The errors of model measurements scale like \req{epemmeas}
derived for $e^+e^-$ collisions.
\section{Conclusion}
We have discussed estimates of $Z'$ exclusion limits and of errors of
$Z'$ model measurements at different colliders.
Simple formulae for the scaling of these limits with the c.m. energy
and the integrated luminosity are given. 
They allow a prediction of the influence of systematic errors on
$Z'$ limits.

The $Z'$ limits at hadron colliders come from direct production,
while the indirect $Z'$ limits at $e^+e^-$ and $e^\pm p$ colliders are
due to interferences of the virtual $Z'$ exchange with the SM
amplitudes.
Therefore, the scaling of $Z'$ exclusion limits and measurements
with the c.m. energy and with the integrated luminosity is the same in 
$e^+e^-$ and $e^\pm p$ collisions.
This implies the same sensitivity to systematic errors.
The limits from hadron collisions scale differently.
Nevertheless, the sensitivity of model measurements to systematic
errors is the same for all colliders. 
$Z'$ exclusion limits are always much less sensitive
to systematic errors than $Z'$ model measurements.

\end{document}